\documentstyle[aps,epsfig]{revtex}

\begin{document}

\twocolumn[\hsize\textwidth\columnwidth\hsize\csname@twocolumnfalse\endcsname
\title{Molecular
Dynamics Simulations of Dynamic Force Microscopy:
Applications to the Si(111)-7x7 Surface }

\author{A. Abdurixit\cite{abdu}, A. Baratoff and E. Meyer}

\address{\emph{Department of Physics and Astronomy, University of Basel,
Klingelbergstr. 82, 4056 Basel, Switzerland}}

\date{Accepted in Applied Surface Science}

\maketitle
\draft

\begin{abstract}

Molecular dynamics simulations have been performed to understand true
atomic resolution, which has been observed on the Si(111)-7$\times$7
surface by dynamic force microscopy in ultra high vacuum(UHV). Stable
atomic-scale contrast is reproduced in simulations at constant mean height
above a critical tip-sample separation when monitoring the interaction
force between tip and sample. Missing or additional adatoms can be
recognized in such scans, although they are less well resolved than native
adatoms. The resonance frequency shift, as well as arbitrary scans, e.g.
at constant force can be computed from a series of force-distance
characteristics. By means of dynamic simulations we show how energy losses
induced by interaction with an oscillating tip can be monitored and that
they occur even in the non-contact range.  
\end{abstract}
\vspace*{1cm}

]

\section{Introduction}

Atomic force microscopy (AFM) has become an important method in
fundamental and applied surface science able to obtain local information
about surface structure and interactions, especially on nonconducting
samples. In this method the surface is probed by measuring the interaction
with a sharp tip or changes produced by that interaction. Atomic
resolution can be obtained if the tip can approach the surface in the
attractive regime without making contact. In such non-contact experiments
it is difficult to measure this interaction force accurately while
avoiding cantilever jump to contact.  However, because the resonance
frequency of the cantilever changes locally if the tip starts to be
attracted to specific surface sites, it became possible to image surface
atoms by maintaining a constant frequency shift $\Delta f$ via feedback
control of the cantilever perpendicular
displacement~\cite{Giessibl,Kitamura}. This shift can be measured very
accurately using frequency demodulation~\cite{Albrecht}. Jump to
contact is avoided by using a large cantilever force constant $k$ and a
large tip oscillation amplitude $A$, such that the restoring force $kA$
considerably exceeds the attractive force at the distance of closest
approach. Under this condition, the tip-sample interaction weakly perturbs
the harmonic motion of the tip, so that $\Delta f$ can be calculated from
the force-distance dependence at any particular site, as first pointed out
by Giessibl~\cite{Giessibl2}

Following the pioneering experiments~\cite{Giessibl,Kitamura}, many
investigators have obtained atomically-resolved images of the
Si(111)-7$\times$7 surface in ultra high vacuum (UHV) using
large-amplitude dynamic force microscopy (DFM) under various feedback
conditions~\cite{Luethi,Guethner,Erlandsson,Luethi2,Nakagiri,Uchihashi}.  
In particular L\"uthi \emph{et al}~\cite{Luethi2} recorded simultaneous
images of the topography, of the time-averaged tunneling current and of
the cantilever excitation, i.e. the damping, at constant $A$ and $\Delta
f$. All images showed characteristic features (adatoms and corner holes)
of the dimer-adatom-stacking fault (DAS) model of the 7$\times$7
reconstruction~\cite{Si(111)-7x7}. Point defects, e.g. missing adatoms
appearing at the same locations in all three images; this convincingly
proved true atomic resolution. Surprisingly, the contrast in the damping
appeared inverted with respect to the other two images.  Other issues are
whether $\Delta f$ jumps at the onset of covalent bonding between dangling
bonds at the tip apex and on adatoms of the sample~\cite{Uchihashi}, and
whether optimum atomic-scale resolution is obtained at a closer
separation~\cite{Uchihashi}, in a soft intermittent contact
mode~\cite{Erlandsson} or even in the range where $\Delta f$ increases
with decreasing tip sample separation~\cite{Guethner}. In this
contribution we show how relevant information about those issues and
quantities measured in large-amplitude DFM can be effectively extracted
from classical molecular dynamics simulations despite the enormous
discrepancy in time scales between the tip oscillation and atomic motions.
 
The rest of the paper is organized as follows: the next section describes
the model and computational details; results are presented in the
following two sections; conclusions are summarized at the end.

\section{The Model}

The Si(111) sample is represented by a slab of eight layers with atoms
coupled via the well-known short-range Stillinger-Weber (SW)
potential~\cite{SW} which, besides a pair interaction, contains a
three-body term favoring tetrahedral covalent bonding. Periodic boundary
conditions are applied laterally to a nearly square supercell encompassing
four 7$\times$7 unit cells. Atoms in the two bottom layers are fixed in
bulk-like positions. Atoms in the following three layers are allowed to
move, but their average kinetic energy is controlled by a thermostat. The
Si(111)-7$\times$7 reconstruction was initiated by removing and shifting
atoms in the next two layers, and placing adatoms in the top layer so as
to approximately obtain the geometry of the dimer-adatom-stacking
fault(DAS) model~\cite{Si(111)-7x7}.  Dimers spontaneously formed between
faulted and unfaulted half-cells in the third layer and adatoms
equilibrated 2.7\AA~ away from their second layer neighbors shortly after
molecular dynamics was started. Subsequent simulations were initiated with
this equilibrated structure.

Focusing on short-range covalent interactions, which presumably give rise
to atomic-scale contrast in the non-contact regime, we model the tip by a
sharp pyramidal silicon cluster with 34 atoms in 6 layers. Like AFM tips
etched out of Si, it has a [001] axis. They were kept at bulk-like
positions in order to isolate effects due to sample deformation and
dynamics.  Because the SW potential vanishes beyond 3.8\AA~, the
longer-ranged pair part of a potential~\cite{BH} which was fit to a
first-principles computation of the interaction between Si layers was
added when considering the interaction between the tip and sample atoms.
This potential has a decay length $\lambda$=0.8\AA~ close to that found
in similar recent calculations~\cite{Perez} of the interaction between a
Si$_{10}$ cluster tip and a Si(111) surface. Compared to Ref.14, the
force $F_z$ computed for the same Si$_{10}$ tip above an adatom using our
interaction has a similar tail, but reaches a minimum (maximum attraction)
at a distance $z=$3.0\AA~ from the unperturbed top layer which is too
strong (-7 nN vs. -2.25 nN). Similar values are obtained with our model
tip. In the non-contact range, our results are thus at least
semiquantitatively correct.
 
The coordinates of the all moving atoms are updated according Newton's
law, using a 5$^{th}$-order Gear predictor-corrector
algorithm~\cite{Allen} with a time step of 0.4fs as advocated by
Stillinger and Weber~\cite{SW}.  The average temperature is controlled by
a first-order feedback loop~\cite{Allen}. The thermalization time
$\tau_{T}$ can be chosen such that the coupling to the thermostat is weak,
but the discrete vibrational spectrum of our finite system is broadened
without appreciable distortions. Because the SW potential led to
appreciable surface diffusion at room temperature (which is not
realistic), the setpoint temperature was fixed at 100K to ensure only
small oscillations about the 7$\times$7 equilibrium structure. Another
important concern is the discrepancy between practical times for
simulations and relevant ones in DFM experiments, e.g. the tip oscillation
period $1/f$ (few ps vs. few ms). Because $\Delta f$ and the
distance-dependent contribution to the damping~\cite{Luethi2}~ are induced
by the tip-sample interaction, the cantilever dynamics \emph{per se} can
be ignored if only its stationary oscillation on resonance is considered.

In order to track tip-induced changes, in each simulation we recorded
snapshots of atomic positions, as well as the time evolution of the
following quantities: the center-of-mass of all moving atoms, their
kinetic energy(temperature), the coordinates of the tip apex and of
selected adatoms, all components of forces on the tip and on those
adatoms, the work done on the tip and the heat transfer to the thermostat. 

\section{Quasistatic scans}

According to perturbation theory~\cite{Giessibl2,baratoff}, a
normalized frequency shift independent of $k$ can be computed in
the non-contact range from the z-dependence of $F_z$ (including possible
reversible deformations),

\begin{equation}
kA\frac{\Delta f}{f_0} = - f_0\int_0^{\frac{1}{f_0}}F_z[z(t)]cos\omega tf_0dt
\label{Eq1}
\end{equation}
where $\omega=2\pi f_0$, $z=d+A(1+cos\omega t)$, $d$ is the
minimum tip-surface distance. In the large amplitude limit $A>>d$,
constant $\Delta f$ approximately corresponds to constant $F_z(d)$ at the 
turning point $z = d$. Quasistatic force-distance characteristics were
obtained by approaching the tip at a constant velocity 
$v_z << v_c = \lambda f_{max}\simeq 1000m/s$, $f_{max}=16THz$ being the
highest
natural frequency of our system (adatom vibration against neighboring
atoms). Such approach curves are essentially $v_z$-independent and
reversible down to a site-dependent separation $z_c$ (e.g. 3.75\AA~ above
adatoms) but strongly $v_z$-dependent and hysteretic below because
irreversible symmetry-lowering jumps of the nearest adatom(s) are then
induced, leading to a transverse force in the nN range. Although our
interaction overemphasizes such instabilities, we expect them to develop
below the inflection point in $F_z(z)$, even before the net attraction
reaches a maximum. Indeed, this inflection signals the appearance of
repulsive forces on the nearest adatoms which can be efficiently released
by sideways displacement. 

\begin{figure}[phbt]
\begin{center}
\epsfig{figure=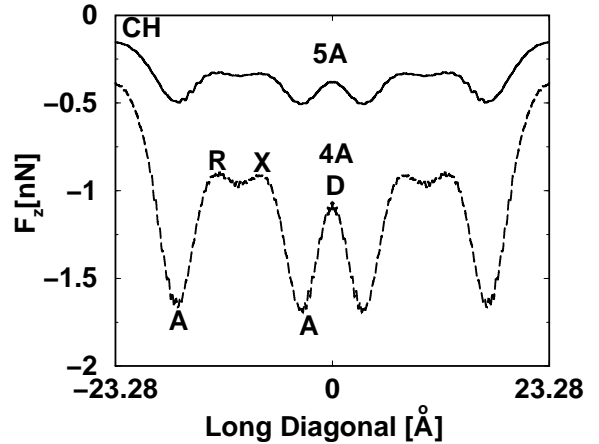, width=7.6cm}
\caption{Variation of the normal force on the tip along the long diagonal 
of the Si(111)-7$\times$7 unit cell recorded at 2 \AA/ps at a height z = 5
and 4
\AA~ measured from the unperturbed adatom layer. The symbols CH, A, R, X, 
D denote the positions of the corner hole, corner and center adatoms,
rest atom, hollow site and dimer row(short diagonal) of the underlying
reconstructed surface} 
\label{4+5} 
\end{center} 
\end{figure}

Below the highest $z_c$ constant height scans recorded at constant
velocity exhibit friction and wear, adatoms being successively picked up
and sometimes redeposited by the moving tip. Above that critical height
such scans are reversible and, as shown in Fig.~\ref{4+5}, exhibit a
contrast with site-specific features in $F_z$ which becomes more
pronounced at smaller $z$. Whereas maximum attraction above adatoms is
expected~\cite{Luethi,Perez}, note that $F_z$ is the same above the rest
atom (R) and above the hollow site (X) in the adjacent triangle between
adatoms, although a recessed dangling bond exists only at the R-site. This
implies that attraction to the adatoms dominates the picture. Because our
model takes no account of charge transfer effects~\cite{Perez}, both cell
halves and all adatoms appear equivalent.

\begin{figure}[phbt]
\begin{center}
\epsfig{figure=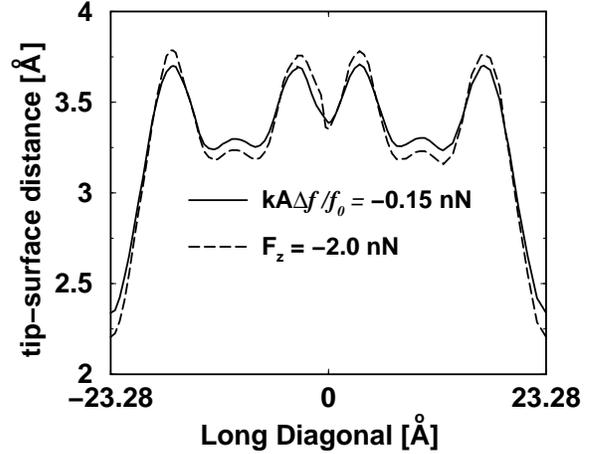, width=7.6cm}
\caption{Topographical scans at constant
force (dashed curve) and constant normalized frequency shift (full
curve) recorded under the same conditions as Fig.
1. Both curves were computed by
interpolation from quasistatic force-distance
characteristics.}
\label{force+freq} 
\end{center} 
\end{figure}

Force-distance characteristics recorded at closely spaced points along a
particular line can be used to reconstruct scans measurable under all
possible feedback operation modes. We checked that the constant height
scan at $z$=4~\AA~ in Fig.~\ref{4+5} essentially coincides with the
corresponding scan reconstructed in this fashion. The same procedure can
be used to reconstruct scans at constant normalized frequency shift using
Eq.(\ref{Eq1}) with the time variable eliminated in favor of
$\tilde{z}=Acos\omega t$ in order to take advantage of the dense set of
data points recorded in approach simulations,

\begin{equation}
kA\frac{\Delta f}{f_{0}}=
-\frac{1}{\pi A}
\int_{-A}^A\frac{F_z(d+A+\tilde z)\tilde z}{\sqrt{A^{2}-\tilde
z^{2}}}d\tilde z
\label{Eq2}
\end{equation}
The rhs of Eq.(\ref{Eq2}) is independent of spring constant $k$ and
oscillation frequency. From the reconstructed topographical scans along
the long diagonal shown in Fig.~\ref{force+freq} we see that the constant
frequency shift scan is close to the constant force curve. The contrast of
the former is slightly reduced due to the non-uniform weight multiplying
$F_z$ in rhs of Eq.(\ref{Eq2}). As expected, the overall contrast is
inverted with respect to the constant height scans in Fig.~\ref{4+5}.
Furthermore, the apparent height difference between corner hole and
adatom sites is approximately 1.7\AA~, i.e. larger than in measurements
at constant $\Delta f$, except one~\cite{Uchihashi}. This is likely due to
our neglect of interactions other than covalent bonding. The contrast we
observe is the maximum possible one for a given normalized frequency
shift or force; the inclusion of Van der Waals and long range
electrostatic interactions would reduce the contrast. The main advantage
of constant height scans is that the contrast should be independent of
such long-range interactions, because they are mainly determined by many
atoms away from the tip apex.

\begin{figure}[phbt]
\begin{center}
\epsfig{figure=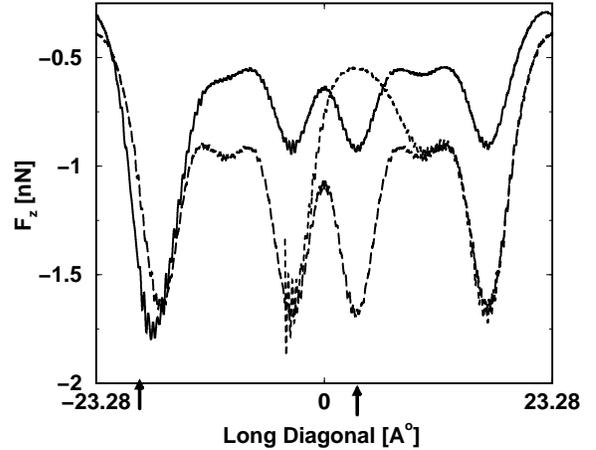, width=7.6cm}
\caption{Constant height scans recorded under the same conditions as Fig.  
1 in the presence of a missing center adatom (short-dashed curve, z=4~\AA),
an additional adatom bound to a corner adatom (full curve,
z=4.5~\AA) compared a similar scan above the perfect surface (dashed
curve, z=4~\AA). The position of the missing (additional) adatom is
indicated by the right (left) arrow on the horizontal axis. The
oscillations appearing at the minima above adatoms are induced by the scan
onset or by thermal fluctuations.}

\label{extra+missing}
\end{center} \end{figure}

In order to assess the resolution of native surface defects, we first
created representative ones by (i) removing a center adatom and (ii)
placing an additional one at one of lowest energy surface sites
$H_3$~\cite{Kaxiras} which is next to a corner adatom, and then relaxed
the
sample. Afterwards we recorded the constant height scans plotted in
Fig.~\ref{extra+missing} together with the scan above
the perfect surface already shown in Fig.~\ref{4+5}. Compared to the
native adatom sites, we observe an almost twice as strong attraction above
the dimer formed between the additional adatom and the nearby corner
adatom, but it is not possible to differentiate the two partners. On the
other hand, the missing adatom produces a maximum almost as high and 
slightly broader than the corner hole. This appearance is consistent with
that of deep minima associated with a missing adatom in constant $\Delta
f$ images~\cite{Luethi2}.

\vspace*{-0.6cm}

\section{Periodic tip oscillation}

\begin{figure}[phbt]
\begin{center}
\epsfig{figure=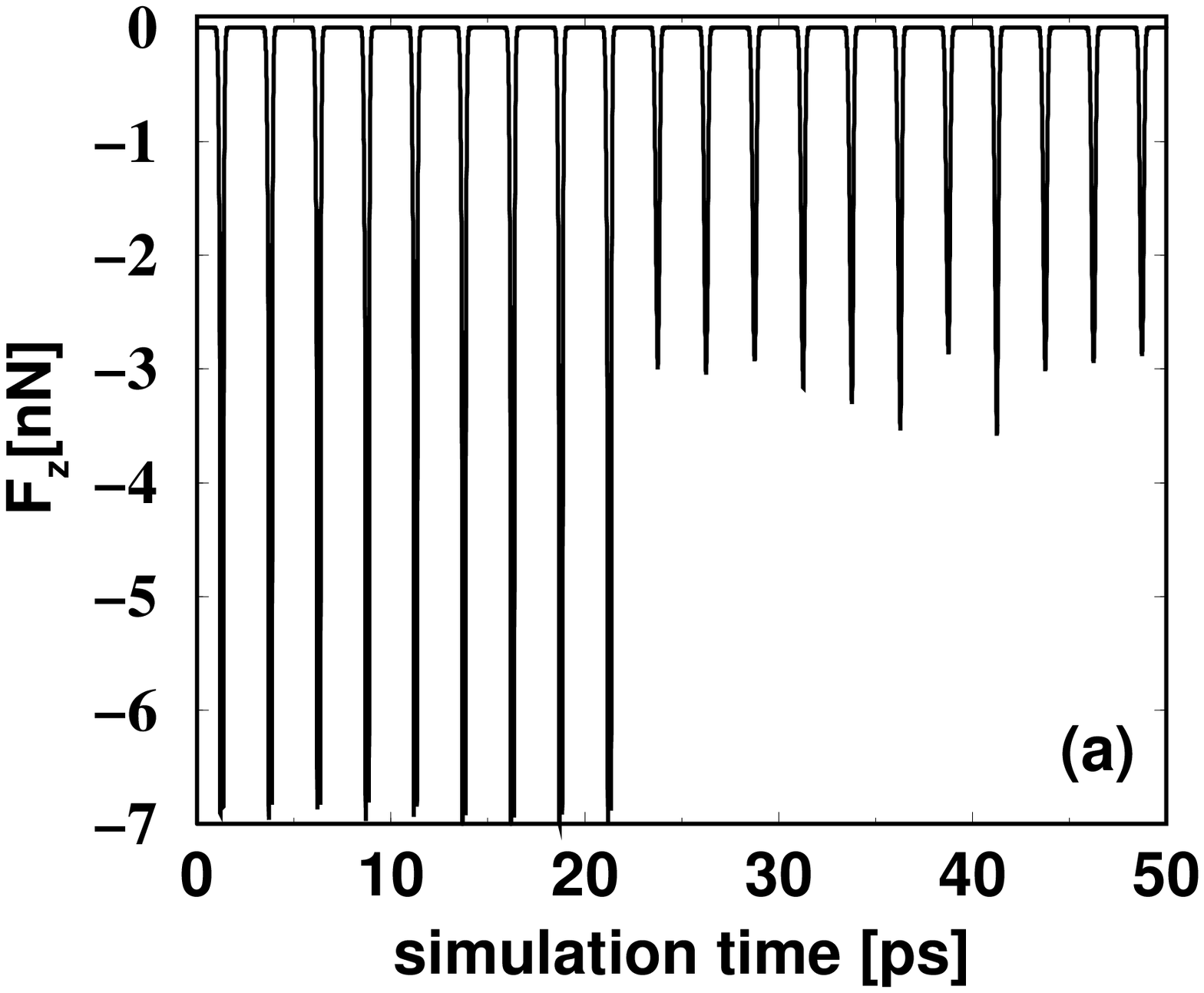, width=7.6cm, height=5cm}
\hskip -0.3cm \epsfig{figure=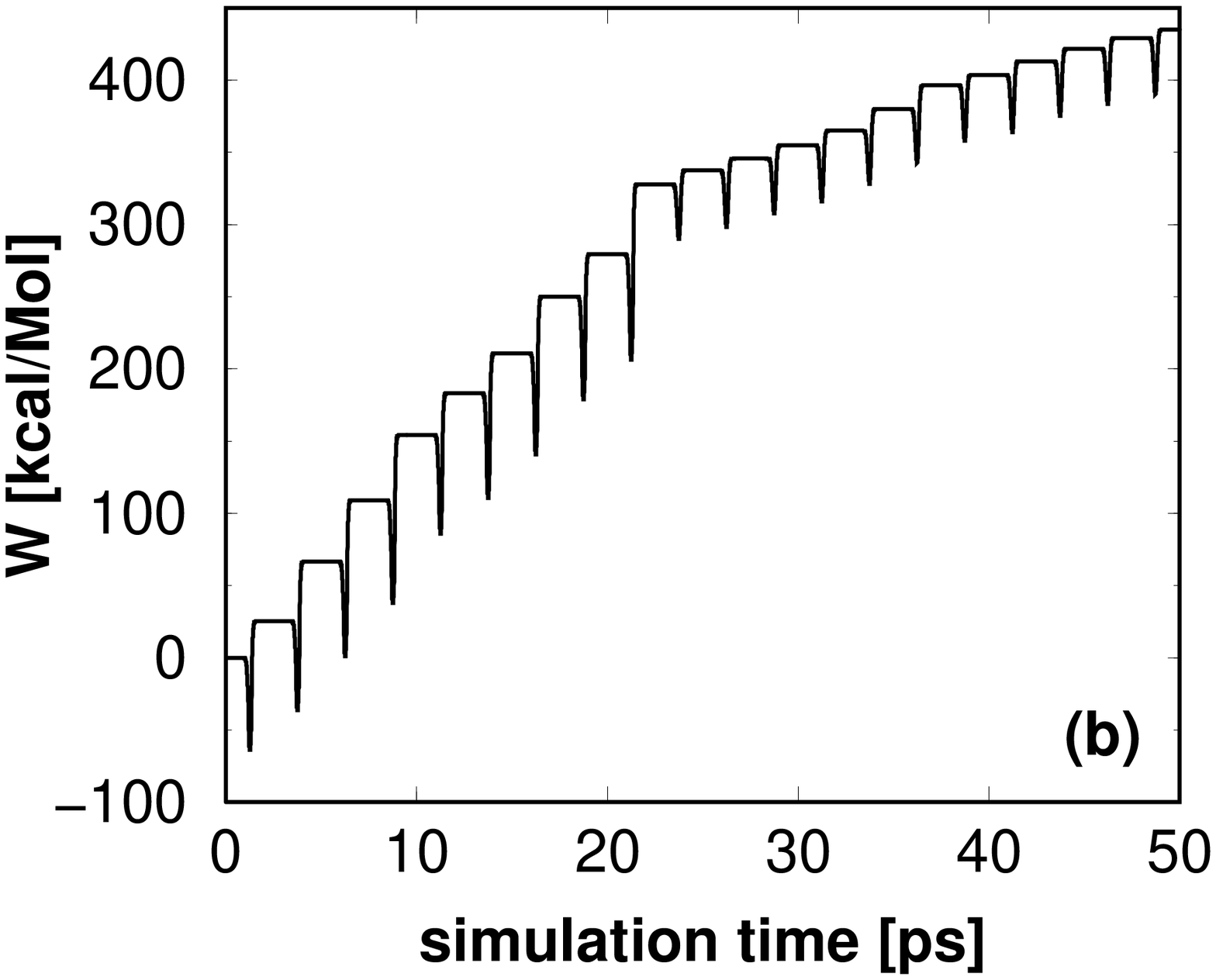, width=7.9cm, height=5cm}
\caption{Top: Time-dependent normal force on the tip oscillated at 0.4
Thz with an amplitude A = 20 \AA~ above a center adatom to a minimum
distance 3.15 \AA~ from the unperturbed adatom layer. Bottom:
Corresponding evolution of the work on the tip; the energy transferred
to the sample after each pulse is converted in to heat(see text).}
\label{force+work}

\end{center}
\end{figure}

In order to determine the distance-dependent damping attributed to the
excitation
of phonons~\cite{Luethi2}, we have performed dynamical simulations in
which
the tip was sinusoidally driven so that it comes within a few angstroms
of the sample. As expected, the time-dependent force on the tip consist of
a periodic series of narrow pulses of width $\sqrt{\lambda/A}/f$ slightly
modulated by thermal fluctuations. Fig.~\ref{force+work} illustrates an
extreme case where this minimum distance is smaller than $z_c$, so that
the center adatom under the tip would be destabilized in a quasistatic
scan. In view of the(unrealistically) narrow pulse width, a jump to an
energetically more favored position on the side of the original site  
occurs only after the ninth pulse. Under typical experimental conditions
the pulse width would be $\approx 1\mu s$ i.e. encompassing
more than one million adatom vibrations, a quasistatic situation.
Nevertheless, a small amount of energy
can be transferred to acoustic phonons, either directly or via anharmonic
decay of local vibrations excited by the force pulses. In our simulations
interaction-induced damping due to both processess~\cite{Luethi2} can be
obtained if the inverse interaction time
$f\sqrt{A/\lambda}$ lies between $f_{max}$ and the frequency of the lowest
mechanical
resonance of our slab (about 1 $THz$). Thus that the imposed frequency
cannot be much below that lower limit.

To visualize energy transfer we calculate the work done by the tip as $W =
- \int F_z v_z dt$. From Fig.~\ref{force+work} we see that apart from
quasi-reversible spikes accompanying the force pulses, $W$ exhibits an
irreversible increase, which comes about because sample atoms do not quite
respond adiabatically. This results in a slight asymmetry of the force
pulses. This irreversible part of the work goes into vibrational energy
which is then shared between different modes i.e. turns into heat. This
transformation is somewhat assisted by our weakly coupled feedback loop
which regulates the total kinetic energy of all thermalized atoms
according to the equipartition rule.

The thermostat time constant $\tau_T$ can be chosen shorter than the
period so that equilibration is practically achieved between successive
pulses. The heat transferred to the thermostat then looks like a staircase
as a function of time, and the average transfer rate coincides with the
mean power transferred by the tip (overall energy conservation) as
demonstrated elsewhere~\cite{abdurixit}.  In those earlier simulations, we
observed that for a given force at the minimum distance d the energy
transfer rate was stronger at the CH site than at the X site, in apparent
agreement with the experimentally observed inverted contrast in the
damping. As expected the damping contrast was in accordance with the
topography if the tip was driven with the same amplitude, but coming
within the same minimum distance d=3.4~\AA. The observation of a contrast
in the damping~\cite{Luethi2} at a constant force $F_z(d)$ is a delicate
issue because such a contrast would vanish if the substrate were treated
as a continuum subject to delta function like pulses~\cite{Duerig}. In
recent dynamic simulations we managed to detect a damping roughly
proportional to $F_z^2(d)$ in the non-contact range, as expected for a
linear response. We found that it is maximum above adatoms, i.e. at sites
where the distribution of $F_z$ among nearby atoms is most localized
laterally.

\section{Conclusions and Acknowledgments}

Molecular dynamics simulations can be conducted and analyzed effectively
to provide valuable information unavailable in experiments, as well as to
compute measurable properties (resonance frequency shift and damping) and
corresponding scan lines. This has been illustrated for the
Si(111)-7$\times$7 surface. Bearing in mind that our model tip is
extremely sharp and that long-range interactions are ignored, the
agreement with experimental constant $\Delta f$ scans is quite
satisfactory. As with any simulations, some details are sensitive to the
underlying model. The interactions we assumed tend to unduly favor adatom
jumps along the surface and to the tip below a separation $z_c$ which
appears somewhat large. Nevertheless, such a critical distance is expected
to appear before the net force reaches a minimum (maximum attraction) upon
approach. It is therefore still difficult to understand how stable DFM
images at constant $\Delta f < 0$ can be obtained following a jump in
$\Delta f$~\cite{Uchihashi} or in the range where $\Delta f$ increases
with decreasing tip-sample separation~\cite{Guethner}. The inverted
contrast sometimes observed in the topography~\cite{Guethner} and in the
damping~\cite{Luethi2} also remains unexplained. New attempts to reproduce
such measurements, with careful attention to possible feedback artifacts
and to the relevant range of setpoints are therefore very desirable. More
work is also needed to understand the detailed dissipation pathways and to
extrapolate the computed damping for comparison with measured values.

The authors thank Prof. H.-J. G\"untherodt and the Swiss National
Research program NFP36 "Nanosciences" for encouragement and support, T.
Bonner for initiating this simulation and his molecular dynamics code, and
M. Bammerlin, R. Bennewitz, M. Guggisberg and R. L\"uthi for helpful
discussions.


\begin{thebibliography}{10}

\bibitem[*]{abdu}Corresponding
author, email: abduxukur.abdurixit@unibas.ch, fax: +41-61-267 37 84/95,
tel: +41-61-267 37 61

\bibitem{Giessibl}F.J. Giessibl, Science 267 (1995) 68

\bibitem{Kitamura}S. Kitamura and H. Iwatsuki, Jpn. J. Appl. Phys. 34
(1995) L145

\bibitem{Albrecht}T.R. Albrecht, P. Gr\"utter, D. Horne and D. Rugar, J.
App. Phys. 69 (1991) 668

\bibitem{Giessibl2} F.J. Giessibl, Phys. Rev. B, 56 (1997) 16010

\bibitem{Luethi}R. L\"uthi, E. Meyer, M. Bammerlin, A. Baratoff, T. Lehmann, L. Howald, Ch. Gerber,
and H.-J. G\"untherodt, Z. Phys. B 100 (1996) 165

\bibitem{Guethner}P. G\"uthner, J. Vac. Sci. Technol. B 14 (1996) 2428

\bibitem{Erlandsson}R. Erlandsson, L. Olsson and P. Martensson Phys Rev. B
54 (1996) 8309

\bibitem{Luethi2}R. L\"uthi, E. Meyer, M. Bammerlin, A. Baratoff, L. Howald, Ch. Gerber, and H.-J.
G\"untherodt, Surf. Rev. Lett. 4 (1997) 1025. The image in Fig.
3(b) of that conference report should represent the topography at a
constant frequency shift, it was unfortunately inverted and mislabelled.

\bibitem{Nakagiri}N. Nakagiri, M. Suzuki, K. Obiguchi and M. Sugimura,
Surf. Sci. 373 (1997) L329

\bibitem{Uchihashi}T. Uchihashi, Y. Sugawara, T. Tsukomoto, M. Ohta, S. Morita, M. Suzuki, Phys
Rev. B 56 (1997) 9834

\bibitem{Si(111)-7x7}K. Takayanagi,  Y. Tanishiro, M. Takahashi, and S.
Takahashi, J. Vac. Sci. Tech. A 3 (1985) 1502

\bibitem{SW}F. H. Stillinger and T. A. Weber, Phys. Rev. B 31 (1985) 5262 

\bibitem{BH}R. Biswas and D. R. Hamann, Phys. Rev. Lett. 55
(1985) 5262

\bibitem{Perez}R. Perez, I. Stich, M.C. Payne and K. Terakura, Phys. Rev.
B 58 (1998) 10835

\bibitem{Allen}M. P. Allen and D. J. Tildesley, Computer Simulation of
Liquids (Oxford University Press, 1989)

\bibitem{baratoff}A. Baratoff and A. Abdurixit, presented at NC-AFM 98
and in preparation

\bibitem{Kaxiras}K. Cho and E. Kaxiras, Europhy. Lett. 39 (1997) 287

\bibitem{abdurixit} A. Abdurixit, T. Bonner, A. Baratoff and E. Meyer,
presented at STM'99 and NC-AFM 99.

\bibitem{Duerig}U. D\"urig, Surf. Interface Anal. 27 (1999) 467

\end{thebibliography}
\end{document}